**Running head: CHANGING MINDS ABOUT ELECTRIC CARS**

# Changing minds about electric cars:
## An empirically grounded agent-based modeling approach

Ingo Wolf[1,2*], Tobias Schröder[2], Jochen Neumann[1], Gerhard de Haan[1]

[1] Freie Universität Berlin, Germany

[2] Potsdam University of Applied Sciences, Germany

\* To whom correspondence should be addressed: Institut Futur, Arnimallee 9, 14195 Berlin, Germany, Phone +49 30 838 52515; Fax +49 30 838 75494; E-Mail ingolupo@gmx.de

**Manuscript under peer review. Do not cite or circulate without authors' permission.**

**Abstract**

The diffusion of electric vehicles (EVs) is considered an effective policy strategy to meet greenhouse gas reduction targets. For large-scale adoption, however, demand-side oriented policy measures are required, based on consumers' transport needs, values and social norms. We introduce an empirically grounded, spatially explicit, agent-based model, *InnoMind Inno*vation diffusion driven by changing *MIND*s), to simulate the effects of policy interventions and social influence on consumers' transport mode preferences. The agents in this model represent individual consumers. They are calibrated based on empirically derived attributes and characteristics of survey respondents. We model agent decision-making with artificial neural networks that account for the role of emotions in information processing. We present simulations of 4 scenarios for the diffusion of EVs in the city of Berlin, Germany (3 policy scenarios and 1 base case). The results illustrate the varying effectiveness of measures in different market segments and the need for appropriate policies tailored to the heterogeneous needs of different travelers. Moreover, the simulations suggest that introducing an exclusive zone for EVs in the city would accelerate the early-phase diffusion of EVs more effectively than financial incentives only.

# Changing minds about electric cars:
## An empirically grounded agent-based modeling approach

### 1.    Introduction

Electric vehicles (EVs– Plug-in hybrid and battery electric vehicles) are seen as a promising technology to reduce carbon emissions and achieve the transition to more sustainable transport. Comprehensive investment in research and development, e.g. in battery technology, is essential to achieve these goals, but technological development alone will not ensure the large-scale diffusion of such innovations. For successful dissemination of new technologies it is also necessary to address the demand side (e.g. Ozaki & Sevastyanova, 2011; Schuitema, Anable, Skippon, & Kinnear, 2012; Tran, Banister, Bishop, & McCulloch, 2012). To this end, we have developed an agent-based model of consumer perceptions and decisions related to innovation adoption in sustainable transport.

While focused on EVs as a technological innovation, our model also helps to answer questions about broader social innovations; i.e., changes in habits and behavioral patterns related to transport. In particular, increasing the use of public transport, bicycles, and car sharing is considered by some as the more important challenge when it comes to organizing the societal transition to more sustainable transport (e.g., Graham-Rowe, Skippon, Gardner, & Abraham, 2011; Kemp & J., 2004; Köhler et al., 2009; Nykvist & Whitmarsh, 2008). Even more than technology adoption, large-scale changes in behavioral patterns depend on the decisions of individual consumers. Numerous studies in psychology have addressed environmental decision-making at the level of individual minds (e.g., Bamberg, 2006; Collins, 2005; Fujii, 2007; Hunecke, Blobaum, Matthies, & Hoger, 2001; Klöckner & Blöbaum, 2010; Steg, 2005; van der Werff, Steg, & Keizer, 2013), but these studies often neglect the complex interactions with broader societal development and the role of other peoples' experiences and decisions when individuals making decisions.

Agent-based models (ABMs) are considered promising tools to study multi-level interactions between individual behaviors and social dynamics (e.g., Bonabeau, 2002; Epstein & Axtell, 1996; Helbing, 2012). Phenomena at the group or societal level (e.g., innovation diffusion) are treated as emerging from multiple interactions of relatively simple behaviors or decisions at the individual level (e.g., changing attitudes). ABMs have become increasingly popular in studies of innovation diffusion in general and research on environmental innovations like alternative fuel vehicles in particular (e.g., Brown, 2013; Eppstein, Grover, Marshall, & Rizzo, 2011; Higgins, Paevere, Gardner, & Quezada, 2012; Shafiei et al., 2012; Sullivan, Salmeen, & Simon, 2009; Tran, 2012a, 2012b; Zhang, Gensler, & Garcia, 2011).

This work answers recent calls for more psychologically realistic models of decision making in ABMs of innovation and social contagion (Kiesling, Günther, Stummer, & Wakolbinger, 2011; Sobkowicz, 2009; Squazzoni, Jager, & Edmonds, 2013; Sun, 2012). Previous models have formalized social contagion and innovation diffusion based on simplistic rules. Many such models are inspired by epidemiological models, in which agents adopt decisions of others simply if they exceed some previously defined threshold (e.g., Deffuant, Neau, Amblard, & Weisbuch, 2000; Deffuant, 2006; Faber, Valente, & Janssen, 2010; Hegselmann & Krause, 2002). Some work on the incorporation of psychological more plausible rules of decision making has been developed (e.g., Jager, Janssen, Vries, Greef, & Vlek, 2000; Schwarz & Ernst, 2009; Tao Zhang & Nuttall, 2011), mainly following the theoretical framework of the Theory of Planned Behavior (Ajzen, 1991; Fishbein & Ajzen, 2010). However, these approaches fail to consider the importance of human emotions in the diffusion process.

In an attempt to overcome some of these limitations of agent-based models of innovation diffusion, the decision and communication mechanisms implemented in our novel *InnoMind* model (for Innovation Diffusion Through Chaning Minds) are based on recent advances in understanding the role of emotion in human decision-making and communication. InnoMind is a multi-agent extension of Thagard's (2006) HOTCO model (for "HOT COherence"), according to which agents make decisions by maximizing the coherence of their current beliefs and emotions. InnoMind agents are susceptible to beliefs of other agents as well as further external influences (e.g. political measures), as they can adopt new beliefs and emotions (i.e. learn). As a consequence, they may change transport mode decisions over time.

In contrast to previous simulation models of EV diffusion, which mainly have considered rational factors of adoption decisions –such as costs, time and driving range– our model accounts additionally for essential psychological factors influencing the individual intention to adopt EVs (cf. Schuitema et al., 2013). Moreover, as recommended in a recent review of EV-diffusion simulation models (Al-Alawi & Bradley, 2013), our agent-based modeling approach extends previous work by rigorously grounding simulated mental representations of agents and the parameterization of social influence in empirical work.

The contribution of the present research is thus threefold: (1) We provide a novel theoretical framework for modeling innovation diffusion based on cutting-edge cognitive science. (2) We show how rich empirical data can be integrated into such a theoretically motivated multi-agent decision model. (3) We demonstrate how this approach can inform strategic decisions related to EV diffusion, where data for classical analysis (e.g., discrete choice models) is not available yet. In particular, we evaluate the effectiveness of various policy interventions designed to enhance the acceptability of and future uptake of EVs separately for different consumer groups.

The novel ABM, which we describe in the following sections, explains how patterns of belief change and innovation diffusion in social systems emerge from psychological processes such as attitudes, values, emotions, social norms, and identity (e.g. Fishbein &

Ajzen, 2010; Gigerenzer & Goldstein, 1996; Homer-Dixon et al., 2013; Kahnemann, 2011; Loewenstein, Weber, Hsee, & Welch, 2001; Mehrabian & Wetter, 1987; Thagard & Kroon, 2006; Thagard, 2006). The model demonstrates how the current structure of mental representations, psychological needs, and social values creates path dependencies and constraints on future possibilities for social change and transport transitions. Based on state-of-the-art theorizing in cognitive science, and grounded in empirical data from focus groups, a representative survey, and a vignette experiment (Wolf, Hatri, Schröder, Neumann, & de Haan, forthcoming; Wolf, Schröder, Neumann, Hoffmann, & de Haan, forthcoming), the ABM can be used to generate psychologically plausible scenarios for innovation adoption. As a case study, we have focused on the city of Berlin, one of the four regions in Germany under the federal government's "Showcase of Electric Vehicles" initiative (NPE, 2012).

The remainder of this paper is structured as follows. Section 2 deals with the model architecture. We explain mechanisms for individual decision-making based on emotional coherence (Section 2.1), for the flows of information based on homophily in social networks (Section 2.2), and for the change of mental representations based on the communication of facts and emotions (Section 2.3). Section 2.4 summarizes the overall algorithm of our model. Section 3 describes the results of the model validation (Section 3.1), a baseline diffusion scenario for different types of consumers (Section 3.2), and simulations of policy scenarios related to the dissemination of EVs (Section 3.3). Finally, in Section 4, we summarize key findings, discuss limitations and practical implications, and provide suggestions for future research.

## 2. The Agent-Based Model: Design and Methods

In this section, we describe our theory of innovation adoption and its implementation in an agent-based model. This theory follows a more general multi-level approach to the study of belief change in complex social systems (Homer-Dixon et al., 2013). We think that peoples' individual decisions about transport result from maximizing the satisfaction of constraints given by their mental representations, which include emotions, needs, priorities, possible actions, and knowledge about the extent to which the different actions facilitate the needs. This mechanism is called emotional coherence and modeled with localist neural networks capable of processing emotions (Thagard, 2006). The adoption of innovation occurs when people change their mental representations as a result of obtaining new information through communication with others or media campaigns, but this is constrained by the compatibility of the new information with the existing mental representations. The model has a mechanism for specifying which two agents communicate with each other at any time step. This mechanism is based on sociological theorizing about homophily in social networks (e.g., McPherson, Smith-Lovin, & Cook, 2001), predicting that the likelihood of two agents exchanging information is dependent on their similarity along socio-demographic variables. In addition, we take into account geographical proximity and individual sociability for modeling social tie formation. For agent-to-agent communication, we assume in our model two possible mechanisms, in line with dual-process models of persuasion from social psychology (e.g., Chaiken, 1987; Chen & Chaiken, 1999; Petty & Cacioppo, 1986).The first mechanism is "cold" and changes the agents' factual knowledge about contingencies between actions and needs. The second mechanism is "hot" and changes the emotional values attached to the different actions. The following sections elaborate on each of these mechanisms.

## 2.1. Agent Decision-Making: Emotional Coherence

Mental representations can be construed as networks of constraints (Thagard, 2000, 2006). Positive constraints are given by elements that go together. For example, taking the bus facilitates the needs of being environmentally responsible. Negative constraints are given by elements that contradict each other. For example, taking the bus is often incompatible with the need for independence. Emotions carry information on how important specific elements are for the individual. Someone with strong environmental values will feel very positive about being environmentally responsible, but to others, the concept might be neutral or even negative.

Decision-making involves the best possible satisfaction of all the given constraints in parallel by organizing mental representations into a coherent set (Thagard & Millgram, 1995). In the example, an environmentalist might decide to take the bus and come to the conclusion that absolute independence is not so important after all. This process of parallel constraint satisfaction can be modeled with connectionist networks, where the nodes are concepts or propositions, excitatory connections between nodes are positive constraints, and inhibitory connections are negative constraints. Decisions then correspond to stable patterns of activated and inhibited elements after multiple rounds of updating the activations of nodes in parallel according to their incoming connections (e.g., Bechtel & Abrahamsen, 2002; Thagard & Millgram, 1995; Thagard, 2000; for mathematical details, see the Appendix).

Emotions can be modeled within such a network by defining special valence nodes that have excitatory (inhibitory) connections with the nodes representing emotionally positive (negative) concepts (Thagard, 2006). In such a HOTCO network model (for "HOT COherence"; Thagard, 2006), valences influence the activation of concept nodes to account for the crucial role of emotion in decision-making and the fundamental psychological fact that all cognition is biased by motivation (e.g., Damasio, 1994; Kunda, 1990; Loewenstein et al., 2001; Thagard, 2006). HOTCO has been applied to various phenomena such as legal decision-making, political perceptions, or religion (Schröder & Thagard, 2011; Thagard, 2003, 2006). In the present agent-based model, we used HOTCO as the basis for the individual agents' transport decisions.

Of course, environmental consciousness and independence are not the only needs that are relevant to peoples' transport decisions. In order to maximize the empirical plausibility of the HOTCO networks representing individual agents in our model, we conducted two empirical studies – qualitative and quantitative – prior to developing the model, the details of which are described elsewhere (Wolf et al., forthcoming). The first study involved four focus group discussions (*N* = 6-8 each). They provided us with a detailed, in-depth picture of people's needs regarding transport as well as their current cognitive and emotional representations of EVs and other means of travel. The architecture of our agents, which is displayed in Fig. 1, is based on the results of focus group discussions. Eight different transport needs with different emotional values are connected with five different means of transport – a total of 40 facilitation relations for each agent. Green lines represent excitatory connections. For example, the agent in Fig. 1 thinks that using a (internal combustion engine) car (the second box from the left in the bottom row) facilitates his need for independence (the rightmost circle in the top row), while using an electric vehicle (the fourth box in the bottom row) would impede that need due to the expectation that EVs' limited driving range provides not the flexibility he needs for his every day mobility. Analogously, an insufficient charging infrastructure might cause the negative assessment that EVs do not facilitate the need for security for this agent. The agents' overall preference, however, is not driven by these two considerations alone. Rather, as explained above, his decision is a function of coherence across all the constraints (i.e. facilitation relations to all five action options, valence weights of needs, and priorities of needs). The agent in our example tends to refuse the use of other transport options, since they satisfy fewer constraints from its belief structure than the use of an internal combustion engine car.

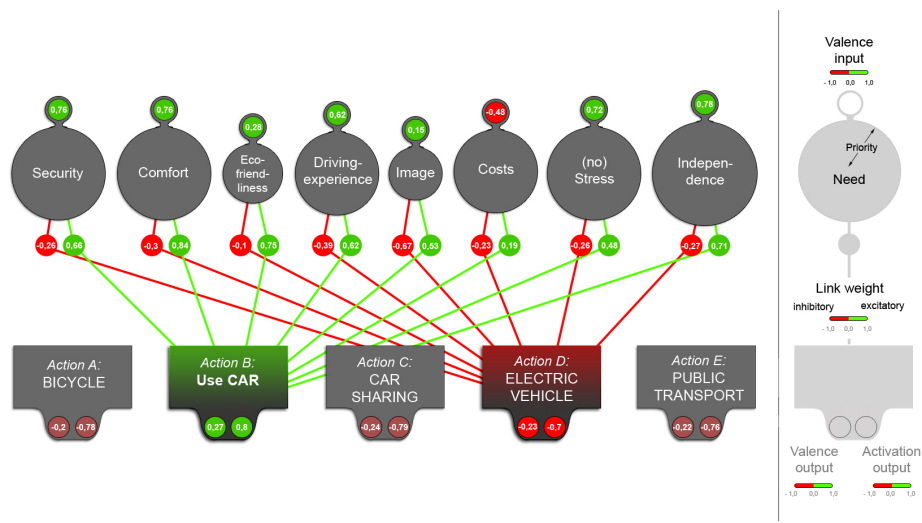

**Fig. 1**. HOTCO emotional constraint network of transport mode decisions. Circle size indicates the priority of a need (large radius = high priority, small radius = low priority). Green (red) lines denote excitatory (inhibitory) connections. Numerical values in the upper and middle green/red circles represent empirically determined weights of emotional and cognitive links. Numerical values in the lower green/red circles denote valences and activations of different action units calculated by HOTCO. Note: For reasons of parsimony, only link weights between two modes of transport and needs are shown. In the agent-based model all five action units are linked to all needs by either inhibitory or excitatory links.

As shown in Fig.1, the agents have three types of input parameters: the priority values and emotional valences of needs and the weights of the links between needs and actions. There are two types of output parameters, the valences and activations of actions. Valences correspond to the emotions associated with the actions, while activations provide a basis for decisions. The action with the highest activation value is assumed to be the most likely one to be chosen (for technical details, see the appendix, or Thagard, 2006). Importantly, a decision in this approach represents a preference for a particular mode of transport rather the actual use of it. Further assumptions and variables in the environment of agents would have to be implemented to formalize this behavior in our model in a realistic manner.

For initializing and calibrating the model, the input parameters of all agents were aligned with empirical data from our second empirical study, a representative survey of the population of Berlin, Germany, with an online questionnaire ($N = 675$) (for details, see Wolf et al., forthcoming). The survey contained rating scales to elicit respondents' appraisals of need importance, valences, and facilitations between needs and actions (e.g., "Driving an electric vehicle offers me the flexibility I need"). For technical reasons, the individual responses were transformed to a parameter range from -1 to +1 and then implemented into the model. Thus, there is a one-to-one correspondence between the survey and the agent-based model; each survey participant is represented as an individual agent.

The agents were tagged according to the classification of their corresponding survey respondents into types of transport consumers. We performed cluster analysis of the original survey data, aimed at creating a parsimonious typology of consumers according to their transport-related needs and attitudes (for details, see Wolf , Schröder, et al., forthcoming). We found that the inhabitants of Berlin can be reasonably described as consisting of four different types, as shown in Table 1. They can be distinguished not only by their values and attitudes, but also by their behaviors (e.g., kms/year driven by car; frequency of taking the bus).

We called type 1 the *Comfort-oriented Individualists*. They have very positive emotions about cars, drive them a lot, and care relatively little about the environment. Their most important transport-related needs are comfort, autonomy, and safety. They are relatively skeptical about electric vehicles. *Cost-Oriented Pragmatics* (type 2) prefer trains and buses, although not for ideological reasons. Their most important needs are cost-efficiency and safety. Among all respondents, they are the least likely to buy an electric vehicle. *Innovation-Oriented Progressives* (type 3) drive a lot, but switch flexibly between cars and public transport. Independence, safety and cost-efficiency are the most important needs for them, when choosing a means of travel. Finally, type 4, the *Eco-Oriented Opinion Leaders* have the highest average level of education. They attach very positive emotions to bicycles, public transport, and electric vehicles.

**Table 1:** Socio-demographic characteristics, travel behavior, preferences and motives of different mobility types

|  | Type 1 | Type II | Type III | Type IV |
|---|---|---|---|---|
|  | Comfort-oriented Individualists | Cost-Oriented Pragmatics | Innovation-Oriented Progressives | Eco-Oriented Opinion Leaders |
| *Socio-demographics* |  |  |  |  |
| % of total sample | 15 | 16 | 34 | 35 |
| Gender (% females) | 55 | 49 | 48 | 46 |
| Education (% graduate population) | 28 | 26 | 24 | 31 |

| Income | | | | |
|---|---|---|---|---|
| < 2500 € p.m. (%) | 76 | 77 | 74 | 72 |
| > 2500 € p.m. (%) | 24 | 23 | 26 | 28 |
| *Self-reported travel* | | | | |
| Kilometers by ICE[1] car | | | | |
| < 15.000 per year (%) | 65 | 89 | 79 | 87 |
| > 15.000 per year (%) | 36 | 11 | 21 | 13 |
| Preferred mode of transport | ICE[1] car | PT[2] | ICE[1] car | PT[2] |
| *Acceptance of EV[3]s* | | | | |
| Intent. to buy an EV[3] (%) | 3 | 3 | 9 | 13 |
| *Motives of mode choice* | | | | |
| Most important needs | 1. Independence | 1. Cost | 1. Independence | 1. Cost |
| | 2. Security | 2. Security | 2. Security | 2. Security |
| | 3. Comfort | 3. Independence | 3. Comfort | 3. Eco-friendliness |

[1] = Internal combustion engine
[2] = Public transport
[3] = Electric vehicle

Please note carefully that tagging the agents with a consumer type has no consequences for the behavior of the agent in the model, since all the agents were calibrated individually. However, we will use the typology below when we describe simulations, to demonstrate how technology adoption dynamics differ across agents with different mental representations to start with.

## 2.2. Social network structure: Who talks to whom?

Empirical research has shown that peoples' attitudes and decisions to adopt new behaviors or technologies are influenced by their social environment and network (e.g., Aral & Walker, 2012; Axsen & Kurani, 2011; Iyengar, Van den Bulte, & Valente, 2010; Valente, 1995, 2005). In our model we generated a social network structure following socio-psychological network models by Hamill et al. (2009), Edmonds (2006), and Mcpherson et al. (1991, 2001) and Blau (1977), given that our previously conducted survey (Wolf, Schröder, et al., forthcoming) did not generate network data. In our combined approach, the likelihood that two agents form a social tie and thus exchange their opinions about transport modes is a function of their geographical proximity, social reach, and socio-demographic similarity (i.e. homophily).

Before creating the interaction structure, we initialized a heterogeneous agent population of 675 agents corresponding to the characteristics of our survey respondents (for an overview of initialized parameters see Table C1 in the appendix). Therefore we used, in addition to cognitive-emotional parameters (see Section 2.1.), individual socio-demographic properties and residential location of our survey respondents, as well as the affiliation to a particular mobility type.

The generation of the social network structure involved three steps. First, each agent was located on a map of Berlin based on the residential location of his real-life counterpart. Since agents do not roam during the simulation, their social reach was determined by a circle surrounding each agent, following Hamill et al. (2009). The radius of the circle comprises a range from 0 to 1 and was grounded empirically on four survey items addressing self-reported opinion-leadership (e.g., "My friends often ask me to give advice upon travel and transport issues") as well as social orientation (e.g., "Before I adopt an innovation, in general I ask the advice of my friends"). Agents with a wide social reach (i.e. radius close to 1) would thus reach more potential communications partners in the geographical neighborhood than those with a circle radius close to 0. But social contacts in this geographical-social environment are not random. Due to the homophilious nature of networks, the probability of an interaction between two agents is a declining function of distance in Blau space, that is, a n-dimensional latent social space (McPherson, 1983). To define social similarity, each agent calculated in a second step the Euclidean distance in a 6-dimensional space for all gents within its social reach. The dimensions of the socio-demographic coordinate system are defined by age, gender, income, level of education, level of modernity, and level of consumption. The location of each agent in social space depends on the characteristics on theses static attributes. This concept of Blau space follows the principle of homophily, according to which the likelihood that two individuals communicate with each other is a function of their socio-demographic similarity (Mcpherson et al., 2001). For mathematical details, see the appendix.

## 2.3. Information exchange between agents: "cold" and "hot" communication

Besides the agents' individual mental representations and the flows of information at the level of the social system, we also modeled a persuasion mechanism that captures belief change as the result of immediate communication. Most psychological theories of information processing and decision-making assume some form of interaction between more deliberate, intentional and more automatic, emotion-driven processes (e.g., Deutsch & Strack, 2006; Kahnemann, 2011; Schröder, Stewart, & Thagard, forthcoming). These "cold" and "hot" aspects of cognition correspond with different variants of theorizing about two different routes to persuasion, central and systematic vs. peripheral and heuristic (e.g., Chaiken, 1987; Chen & Chaiken, 1999; Petty & Cacioppo, 1986). Loosely based on this well-established dichotomy in psychological research, we allow our agents to adapt their mental representations in communication through two different mechanisms, taken from an earlier multi-agent variant of HOTCO (Thagard & Kroon, 2006). In this model, aimed at simulating decision-making in small groups, communication can be about facts (e.g., the information that a certain action will facilitate achieving the agent's needs), and is called the means-ends mechanism. Communication can also be emotional (e.g., expressed enthusiasm or emotional attachment about one action), and is called contagion (cf. Hatfield, Cacioppo, & Rapson, 1993).

In parallel constraint satisfaction models such as Thagard's (2006) HOTCO, belief adjustments in response to external input can be implemented by changing connection weights between elements of the network (e.g., Monroe & Read, 2008; Read & Urada, 2003;

Van Overwalle & Siebler, 2005). When the constraint-satisfaction algorithm described in Section 2.1 and appendix is then applied again, the network might settle in a different stable state than before. This is how we model changes in mental representation that follow agent-to-agent communication. The two communication mechanisms described above impact different sets of connection weights in the receiver neural network. Means-ends communication results in an adjustment of the links between need and action nodes (see Fig. 1). For example, one agent might transfer to the other the factual information that electric vehicles are environmentally friendly, resulting in a stronger excitatory link between the node representing the need for eco-consciousness and the node representing the action of driving an electric vehicle. Communication by emotional contagion results in an adjustment of the valences of action nodes, i.e. the connection weights between the action nodes and a special valence node in the network (for details, see Section 2.1 and the appendix). For example, this valence adjustment models the enthusiasm one agent might express to the other through nonverbal cues while talking about her experience when test-driving an electric vehicle.

The parameterization of the connection weights adjustment was based on data from a separate experimental study, aimed at quantifying how people change their beliefs about EVs in response to influence of others (Wolf, Hatri, et al., forthcoming). The experiment was driven by the hypothesis that the acceptance of others' opinions and the process of belief adjustment is a matter of belief strength, emotional valence, and attitude congruence between the sender and the receiver (cf. Osgood & Tannenbaum, 1955). We studied these coherence effects in a vignette experiment by asking participants to rate their agreement and their perceived belief change on a series of unrelated statements about the use of electric vehicles and combustion engine cars. We used the experimental data to determine for different configurations of senders' belief strength and sender-receiver belief congruence the weight changes that optimized the prediction of the empirical data with our persuasion model. Details are given in the appendix.

## 2.4. Summary of the ABM Algorithm

Fig. 2 summarizes how the agent-based model works. At each time step of a simulation, each agent randomly chooses one communication partner out of his individual social network (Fig. 2a). Of course this a simplification of the communication processes in real life. However, we believe it is sufficiently realistic to model the relevant aspects of social influence in our approach. As in other models of social contagion and opinion diffusion (Castellano, Fortunato, & Loreto, 2009; Deffuant et al., 2000; Hegselmann & Krause, 2002; Sobkowicz, Kaschesky, & Bouchard, 2012) the time scale in our model is abstract and without additional assumptions does not immediately correspond to real time units. The content of communication for each conversation is selective and depends on the sender's belief strength and his strength of emotional reaction to transport mode options. Thus, the speaker communicates only factual arguments (i.e. facilitation weights between need and action units) with high confidence and valence connotations of actions that are affectively rich (Fig. 2b).

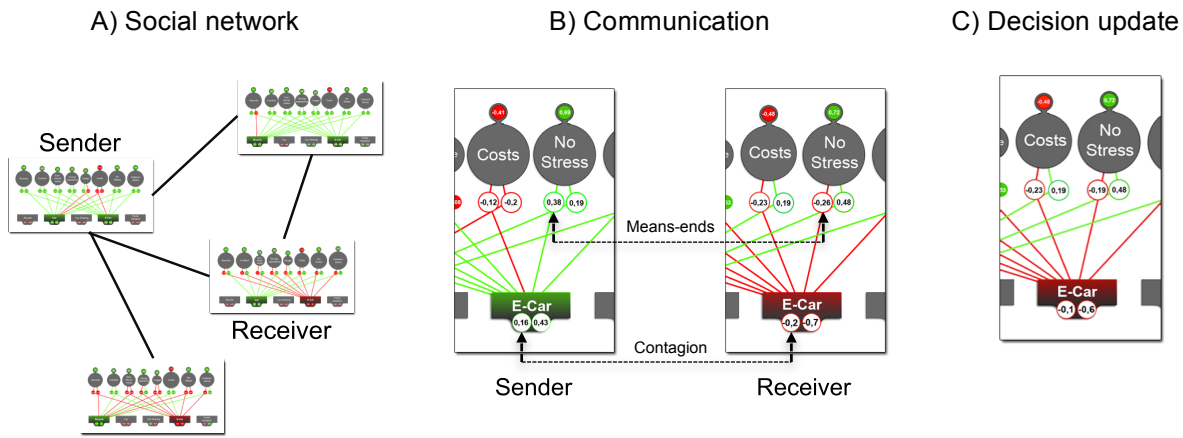

**Fig. 2 a-c**. Illustration of the agent-based model InnoMind. A) An agent randomly selects a communication partner in his personal social network. B) During communication, agents exchange factual and emotional information ("means-ends" and "contagion", respectively. C) Sender and receiver integrate the new information in their individual HOTCO network and recalculate their transportation decisions.

The crucial parts of this communication procedure are the evaluation and integration of information in the mental model of the listening agent. During each conversation, the communication partners assume the roles of speaker and listener simultaneously. Thus, belief adjustment is carried out for both agents of the dyad. Once all affected connection weights of both agents are set to their new values, agents update their decisions employing the coherence algorithm described in Section 2.1 (Fig. 2c). At this stage, they might switch their preferred mode of transport, if the persuasion attempt of the other agent was successful. In the following step of the simulation, agents communicate about transport issues based on their updated mind set with their other social peers. Again, it is important to note that preference changes in our approach represent a change of behavioral intentions rather than a mode switch on a behavioral level.

In addition to peer influence, agents can be influenced by media coverage. To simulate media campaigns about certain policy instruments we implemented a media agent affecting mental representations of agents in a similar manner as in the dyadic communication procedure described above. The media agent has directed links to an adjustable proportion of the agent population (from 0% to 100%) and transmits, at predefined time steps, facilitation weights, priorities, and emotions, depending on the content of the campaign. For example, the information about the introduction of a purchase subsidy for EVs is represented by strengthening the positive facilitation weight between the need unit *Cost* and the action unit *Option D: Electric vehicle*. Analogously to the dyadic communication mechanisms described above, affected agents integrate the information from the media agent in their connectionist network and update their decision. The relevance of received information is adjusted by a policy impact factor, derived from empirical ratings of related items in the survey described above and in Wolf et al. (forthcoming) (e.g. "Would this policy measures change your attitude toward EVs?"). Participants stated their responses on a six-point Likert scale ranging from "Absolute no influence" (1) to "Very strong influence" (6). For technical reasons, these values ware transferred to a parameter range from 0 to 1 (for an overview of mean values and a sensitivity analysis of the policy impact factor see Table C2 and Figure F1 in the appendix).

A formal, mathematical description of the agent-based model is provided in the appendix. The model was implemented in a computer program written in Java. We now turn to a description of its validation and use for simulating the diffusion of EVs under a base case as well as different policy scenarios.

### 3. Results and discussion

### 3.1. Validation of the decision algorithm

Prior to performing a series of simulations, we compared model predictions with data on actual transport choices from the above-mentioned empirical survey (Wolf, Schröder, et al., forthcoming), to validate the accuracy of the connectionist model. To this end, the model computed individual agents' decisions about their preferred transport modes, prior to communication, based on their empirically grounded mental representations (i.e. attitudes, priorities and emotions). Recall that each agent has a real-life counterpart in the empirical survey. For the model validation, survey participants' scores related to self-reported transport behaviors were regressed on the output parameters of the HOTCO networks representing the preferences of these respondents in the agent-based model (i.e., the activation parameters of the action nodes). Note that we calibrated the decision structure of agents exclusively by the variables representing beliefs and emotions. The data on behaviors were used only for validating the model output, but nor for calibrating the model. Social psychological research on the attitude-behavior relationship under the influential Theory of Planned Behavior (Ajzen, 1991; Fishbein & Ajzen, 2010) shows that stated intentions generally account for roughly a third of the variance in behavior (Armitage & Conner, 2001), which we thus considered as a benchmark for the predictive power of our InnoMind model.

We performed stepwise binary logistic regressions to assess the effects of the obtained activations on the probability to use a specific transport mode. Original responses variables on five-point Likert-type scales (ranging from "I use this transport mode (almost) never" (1) to "I use this transport mode (almost) every day" (5)) were dichotomized (level 1-3 into "0 = no use" and level 4 to 5 into "1 = use") and then treated as criteria. We opted for logistic regression and dichotomization because of the non-normal distribution of the error terms in linear regression. In four separate models, we regressed travel mode choice behavior (for EVs, the intention to substitute the current main transport mode) on activations of action units of the decision model (preferences toward internal combustion engine (ICE) car, electric vehicle (EV), public transport (PT) and bicycle (BI)). Again, car sharing was excluded from the analysis due to lack of data. The results for all regression models are shown in Table 2. R square (Nagelkerke) values between 0.231 and 0.462 and percentages of correctly predicted cases varying between 68.6% for model 2 and 78.4% for model 4 indicate in comparison to the power of the established Theory of Planned Behavior (Ajzen, 1991; Fishbein & Ajzen, 2010; see Armitage & Conner, 2001 for a meta-analysis) a reasonable fit and good overall performance of the simulation model. Odds ratios (Exp($B$)) in model 1 (Table 2) indicate that on values for ICE cars have the strongest positive effects on the actual use of ICE cars, while an increase in activations towards PT decreases the probability of car usage. Survey respondents' intentions to substitute the current main mode of transport by an EV (model 2) are significantly predicted by activations of EV action nodes in the corresponding virtual agents. Agents with preferences for public transport are less willing to use EVs, whereas attitudes towards ICE cars and bicycles had no effect on odds of this outcome. The results in model 3, predicting the frequency of public transport use, show again an inverse pattern of two predictors. Activation of PT nodes in virtual agents predicts survey counterparts' use of public transport, while high activations for ICE cars are negatively related to this travel mode use. Model 4 was designed to investigate the use of bicycle and shows high robustness in predicting this travel mode choice (78.4% correct overall; $R^2 = 0.462$). The independent variables attitudes towards EVs and public transport did not have a significant influence on the frequency of use bicycles. The likelihood of use was increased if agents favored bicycle and decreased for car-oriented agents.

**Table 2:** Odds ratio (*Exp(B)*) and their related score statistics (in parenthesis *p*-values) of binary logistic regression analyses with activation values of five different action units as independent variables and the self-reported frequency of transport mode use as dependent variable

| Independent variables | Model 1 Use of internal combustion engine car | Model 2 Intention to substitute current main mode of transport by EV | Model 3 Use of public transport | Model 4 Use of bicycle |
|---|---|---|---|---|
| *HOTCO activations for:* | | | | |
| ICE[1] car | 3.705 (0.001) | n.s. | 0.407 (0.001) | 0.535 (0.001) |
| EV[2] | 1.438 (0.036) | 4.820 (0.001) | n.s. | n.s. |
| PT[3] | 0.524 (0.001) | 0.665 (0.001) | 3.118 (0.001) | n.s. |
| BI[4] | n.s | n.s | n.s | 9.502 (0.001) |
| Constant | 1.394 (0.002) | 1.056 (0.534) | 1.214 (0.040) | 0.610 (0.001) |
| Model chi square | 165.730 | 127.813 | 146.699 | 281.284 |
| Nagelkerke´s $R^2$ | 0.298 | 0.231 | 0.261 | 0.462 |
| -2 log likelihood | 717.529 | 800.453 | 787.804 | 623.0729 |
| % Correct overall | 73.5 | 68.6 | 70.1 | 78.4 |

[1] = Internal combustion engine
[2] = Electric vehicle
[3] = Public transport
[4] = Bicycle
n.s = Not significant

Taken together, these statistical results provide evidence that the parallel constraint satisfaction network model we used is a valid approach to model individual travel mode choice behavior. This conclusion is also upheld by comparing the percentage of internal combustion engine (ICE) cars as the primary means of travel predicted by the model with data on private transport share across the different districts of Berlin, taken from a recent study by the government of Berlin (Senat Administration of Berlin, 2010). A visual comparison is displayed in Fig. 3. Moreover, we used a Kendal rank correlation test to assess the similarity between rank of simulated share of ICE car use across districts and their share in real life. The results indicate ($r = .42$; $p = .055$) that the predictions derived from our model roughly match the observed geospatial pattern of peoples' actual transport decisions–as shown in Fig. 3b. We now turn to scenarios of future EV adoption in Berlin, which we created by simulating communication between the agents in our model.

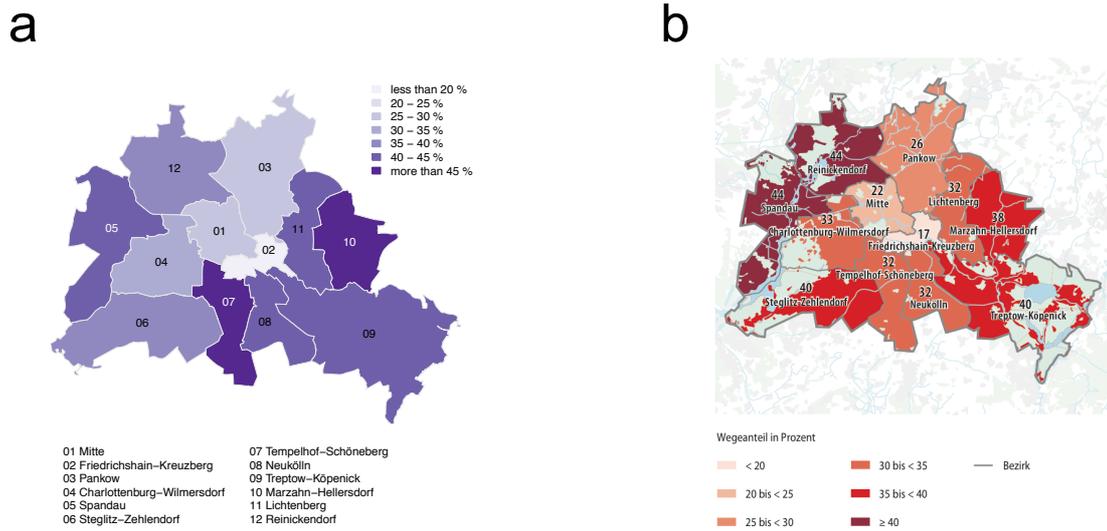

**Fig. 3 a-b.** Districts map of Berlin. A) Percentage of agents preferring internal combustion engine (ICE) car after initialization. Source own illustration. B) Percentage of individual internal combustion engine cars on everyday mode split across the different districts of Berlin. Source: Senat of Berlin.

## 3.2. Scenario descriptions and assumptions

Four scenarios, including a reference scenario, were analyzed to explore the interrelated effects of individual attitude factors (i.e. needs, emotions etc.), social influence, and policy interventions on transport choices in general and on the acceptance of electric vehicles (EVs) in particular. The scenario experiments differ with regard to assumptions we made about external influences, i.e. policy interventions and social influence, on consumers' mental representations. In the baseline scenario, communication with peers alone affected the attitudes of agents. In the policy scenarios, interventions via the media agent additionally affected attitudes toward EVs in specific ways, as described below. Note that the (external) influences of the environment did not directly affect the final decisions of agents, but instead the underlying psychological associations between transport mode options, needs, and emotions. In all scenarios, model parameters of the 675 agents were initialized with the empirical values from our online survey. Each simulation involved 100 time steps[1]. Below, we report average results across 10 runs of the model for each scenario. Drawing on our previously identified typology of consumers, we aggregated the dynamics of consumer decisions separately for the four consumer types, as they substantially differ in their current travel behavior, needs, attitudes and potential to adopt EVs.

As a *Reference case* scenario, we first ran the simulation without any policy intervention. In this scenario, individual attitudes and emotional associations towards transport mode options are influenced only by agent-to-agent communication within the artificial social network, as discussed in Section 2.2. Besides the function as reference for comparison, we used this scenario to identify the susceptibility or resilience of decisions to social influence in a population with highly heterogeneous mode preferences.

Recent policy studies mainly focused on the effects of economic instruments to stimulate the up-take of EVs (Berestenau & Shanjun, 2011; Diamond, 2009; Gallagher & Muehlegger, 2011; Higgins et al., 2012; Sallee, 2008; Shafiei et al., 2012). Our simulation model, however, is designed to evaluate the impact of a wide range of interventions on consumer transport mode decisions that go beyond pure monetary incentives. To illustrate the suitability of our model for policy analysis, we present three separate scenarios assessing the effectiveness of policy instruments to encourage consumers to adopt EVs. The selected measures are meant as examples and not exhaustive for a comprehensive policy analysis. We selected the policies explained below since they are currently the most salient and controversially discussed measures for EVs in the political debate in Germany. The scenarios comprised (a) the establishment of a zero-emission-zone in Berlin, allowing only EVs and other zero emission vehicles to operate in a certain area of the city (*Zero-emission-zone* scenario), (b) the exemption of EVs from motor tax (*Tax exemption* scenario) and (c) the direct subsidy (in the amount of € 5000) towards the purchase of an EV (*Purchase subsidy* scenario). In contrast to the *Reference case*, agents' mental representations and subsequent decisions were not only affected by word of mouth in their social network, but agents moreover considered the consequences of policy measures in decision making. Corresponding to social influence, we therefore assumed that agents' belief-action representations (i.e. facilitation weights between needs and actions units) are modified by the influence of policy-related information in a specific manner for each intervention. To simulate repeated mass media campaigns, we transmitted in all scenarios information about the policies in every tenth time step (i.e. time step 0, 10, 20, 30 etc.) to 70% of the agent population by means of the media agent described above (Section 2.4). After receiving this information, all agents reevaluated their transport mode decisions and carried on with the usual dyadic communication.

---

[1] Since no empirical data about the frequency of communication about transport issues in social peers was available, we were not able to make a valid assumption regarding the time horizon of our simulations. Thus, it is difficult to map the simulation results to real physical time. However, assuming for the sake of argument a basis of 5 conversations about the topic per year, 100 times steps roughly represent a period of 20 years. Nevertheless, we emphasize that the simulations are not intended to be forecasts over a certain time horizon but to compare the effects of different policies on the diffusion process in the early stages.

In the *Zero-emission-zone* scenario we conjecture –in accordance with empirical findings (Wolf et al., forthcoming) – that users would perceive higher levels of independence (i.e., flexibility by being allowed to enter all areas of the city) and a decrease in their stress-level caused by less traffic volume. In the logic of our parallel constraint network model (see Fig. 1), the information about this policy increases the facilitation relations between the needs of "independence" and "no stress" and the action "use EVs". Importantly, the additive changes of individual agent facilitation weights were multiplied by an empirically determined factor of policy impact (for details see Section 2.4 and the appendix).

In both the *Tax exemption* and the *Purchase subsidy* scenarios we modeled the effects of fiscal incentives on widespread acceptance of EVs. Even though these policies affect different aspects of total cost of ownership of EVs -namely purchase versus operating costs- both equally cause in our model a positive shift of facilitation relations between the need of "cost efficiency" and the action "use EVs". In other words, agents believe due to the introduction of these policies that EVs accomplish their need of a cost efficient means of transport to a greater extent. These assumption are supported by empirical evidence showing that consumers do consider different financial benefits associated with alternative fuel vehicles in their adoption decisions (Chandra, Gulati, & Kandlikar, 2010; Diamond, 2009; Rogan, Dennehy, Daly, Howley, & Ó Gallachóir, 2011; Ryan, Ferreira, & Convery, 2009; Sallee, 2008), but are less accurate in distinguishing and estimating the actual economic value of these instruments (cf. Greene, 2010; Larrick & Soll, 2008; Turrentine & Kurani, 2007). Analogous to the *Zero-emission zone* scenario the individual impact of the two policies measures were weighted (i.e. multiplied) differently based on empirical appraisal ratings or our survey participants (e.g. "Would this policy measures change your attitude toward EVs?"). The two scenarios thus have the same structural effect on the agents' belief networks, but they differ in the quantitative strength of this effect.

## 3.3. Simulation results

In this Section, we examine the results of simulating transport mode choices under the different scenarios. First, we discuss the reference case – the dynamics of transport mode preferences through communication among agents alone, without external intervention. Second, we evaluate policies promoting the diffusion of EVs relative to adoption trends in the base case. In both subsections we use the percentage of agents who prefer a particular mode of transport as a measure of impact –we refer to these agents as the "fraction of potential adopters". Recall that preferences are dynamically constructed based on pre-existing mental representations of agents and potentially changed through communication with other agents. These preferences may be interpreted as a mental preparedness to adopt a certain mode of transport. Thus, scenario results should not be interpreted as immediate market predictions, but rather as an explorative approach to investigate the resilience of current mental representations and travel behaviors to external influences.

### 3.3.1 Reference case scenario

The *reference case* scenario captures the influence of social communication on changes in individual travel choices of agents over time, without external intervention. Fig. 4 shows the proportion of users of five transport modes over time, averaged over 10 model runs and separately for the four consumer types described in Section 2.1. Overall, preferences in the heterogeneous agent population remain relatively stable over the 100 time steps and exhibit low fluctuations. The modal split remains substantially different between the different types of travellers. In the group of *Comfort-oriented Individualists* (Fig. 4a) the use of combustion engine (ICE) cars continues to dominate travel choices, with initial shares at $t_1$ of 91% and of 89% by iteration 100. The graphs representing alternative modes indicate that EVs (form $t_1 =$ 3% to $t_{100} =$ 4%), car sharing (from $t_1 =$ 1% to $t_{100} =$ 1%), public transport (from $t_1 =$ 5% to $t_{100}$ = 5%) and bicycles (from $t_1 =$ 0% to $t_{100} =$ 1%) cannot compete with ICE cars in this segment. For the *Cost-oriented Pragmatics* (Fig. 4b) public transport continues to be the most attractive travel mode (36%), followed by bicycles with a slightly decreasing share from 27% ($t_1$) to 23% ($t_{100}$) and a constant subgroup of agents (25%) that favors the use of ICE cars. Although starting from a low level, the preferences for car sharing and EVs increase considerably in this segment by 100% (from $t_1 =$ 5% to $t_{100} =$ 10%) and 60% (from $t_1 =$ 5% to $t_{100} =$ 8%), respectively. *Innovation-oriented Progressives* (Fig. 4c) exhibit a slight decrease in their dominant shares of ICE cars (from $t_1 =$ 49% to $t_{100} =$ 43%) and EVs (from $t_1 =$ 25% to $t_{100} =$ 24%). The simulations show an inverted trend for agents' preferences in this traveller group related to public transport (increase from $t_1 =$ 15% to $t_{100} =$ 17%) and car sharing (increase from $t_1 =$ 2% to $t_{100} =$ 8%). As Fig. 4d illustrates, the almost equally distributed shares of bicycles (27%), public transport (28%) and EV (29%) users in the segment of the *Eco-oriented Opinion leaders* slightly diminish in favor of car sharing (from $t_1$ = 8% to $t_{100}$ = 14%) and ICE cars (from $t_1$ = 7% to $t_{100}$ = 8%).

To sum up, the simulation results indicate that social communication among peers alone causes at most marginal choice shifts in all the four consumer groups. Agents show a high resilience of their overall transport mode decisions, yet a few travelers exhibit the propensity to switch from car use to alternative travel modes (Fig. 4a and Fig. 4b). Interestingly, while car sharing accounts at the beginning of the simulation only for a very low overall share (between 0% and 8%), we observed the most considerable changes of all with regard to this mode option across three groups (Fig. 4b-d). This behavior of our model is consistent with a recently observed surge in the use of car sharing in the city of Berlin (Bock

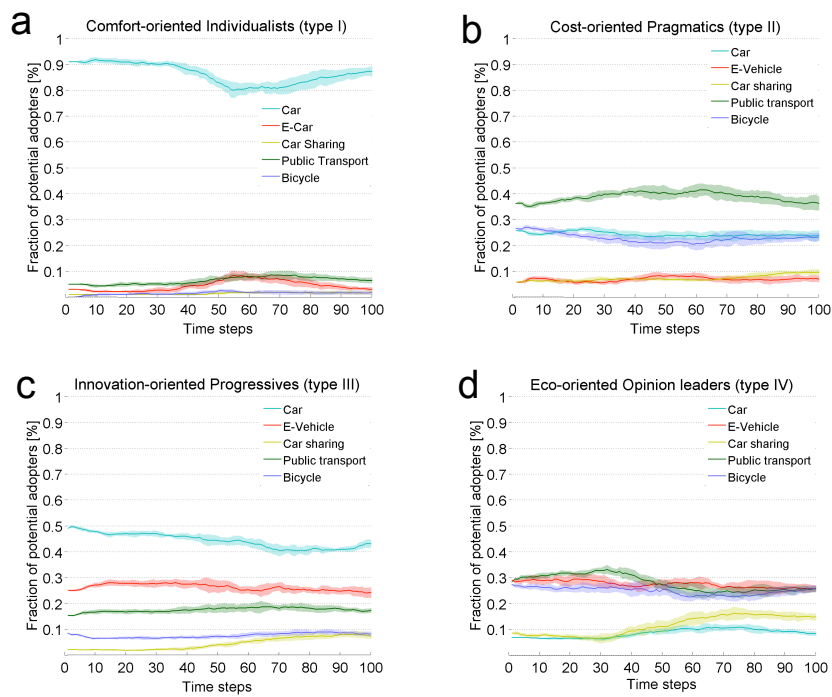

**Fig. 4 a-d.** Diffusion of preferred modes of transport. Modal split over time of (a) *Comfort-oriented Individualist* (mobility type I), (b) *Cost-oriented Pragmatics* (mobility type II), (c) *Innovation-oriented Progressives* (mobility type III) and (d) Eco-oriented Opinion leaders (mobility type IV).

et al., 2013).

### 3.3.2 Electric-vehicle policy scenarios

The results of simulating three policy interventions, designed to accelerate the uptake of EVs, are depicted in Fig. 5 along with the potential adoption rates of EVs from the reference scenario for comparison. Quite plausibly, the simulations suggest that the four consumer types will respond differently to the policy measures (see Fig. 5a-d). *Comfort-oriented Individualists* (Fig. 5a) –the segment exhibiting the lowest EV acceptance rate in the reference case scenario– show increased propensity to adopt EVs in all policy simulations. Nevertheless, ICEs cars still dominate the modal share (between 74% to 77%) in this consumer group. The introduction of the zero-emission-zone leads to a temporary gradual increase of potential EV users, followed by a slight drop (compared to reference case at $t_{100}$ = +13%) below the EV share in the *Purchase subsidy* scenario (+16%) and the *Tax exemption* scenario (+13%).

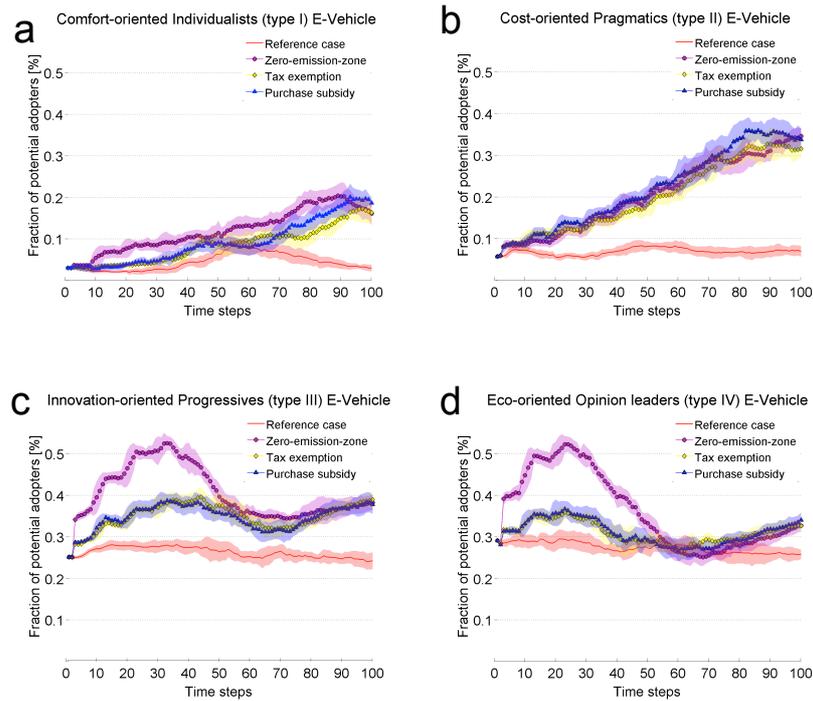

**Fig. 5 a-d.** Diffusion of E-Vehicles in *Reference case*, *Zero-emission-zone*, *Tax exemption* and *Purchase subsidy* scenarios. Panel (a) presents the fraction of potential adopters for *Comfort-oriented Individualist* (mobility type I), panel (b) for *Cost-oriented Pragmatics* (mobility type II), panel (c) for *Innovation-oriented Progressives* (mobility type III) and panel (d) for Eco-oriented Opinion leaders (mobility type IV).

Simulations suggest that the most significant changes triggered by policy measures in the modal share of EVs are to be expected in the sub-population of *Cost-oriented Pragmatics* (Fig. 5b). Both monetary policies increased the fraction of potential EV adopters linearly about 4.9-fold (up to a 34% share) in the *Purchase subsidy* scenario and about 4.4-fold (up to a 31% share) in the *Tax exemptions* scenario. Thereby the purchase subsidy intervention accelerated the acceptance most effectively almost throughout the whole simulated diffusion process. The introduction of a zero-emission-zone resulted in a 5.0-fold (up to 35% share) increase in EV use and thus showed the strongest effect compared to the reference case. The diffusion of EVs in the three policy scenarios comes about as a mode shift predominantly away from public transport (≈ -24%), bicycle (≈ -20%) and ICE cars (≈ -16%).

Fig. 5c and 5d illustrate a high level of fluctuation in EV shares across the simulated policies for the mobility types *Innovation-oriented Progressives* (Fig. 5c) and *Eco-oriented Opinion leaders* (Fig. 5d). Agents of these segments already show relative high acceptance rates of EVs to start with (25% of type III and 29% of type IV). The steep increase of the fraction of potential EV adopters in the *Zero-emission-zone scenario* up to 53% (type III) and 52% (type IV) indicates both that (i) a considerable number of agents in both traveler groups have a tendency to switch their current mode towards EVs and (ii) excluding ICE cars from certain urban areas leads to the most pronounced preference changes in any of the simulated scenarios. However, the following drop demonstrates that neither restrictive nor incentive policies are able to generate sustainable long-term preferences towards EVs at the peak level in these consumer groups. This indicates that consumers potentially willing to adopt EVs rethink their decisions in the light of alternative options. Triggered by the discussion with their social peers, they switch back to modes of transport, which are more commonly known and accepted. Technically, two effects cause this behavior. First, due to non-linear susceptibility to the opinions of others (for details see Appendix D) increasing coherence between the beliefs of agents leads to reinforced mutual influence between communication partners. Thus, if agents with strongly positive attitudes towards at least two transportation modes–e.g. such as *Innovation-oriented Innovators* show for EVs and ICE cars–meet with agents with similar mental representations, this function causes oscillations between both choices since their underlying beliefs are influenced in similar strength. Second, over the course of time the activation values of two or more choices (i.e. preferred mode of transport) can temporarily be identical, thus they are not mutual exclusive.

The zero-emission-zone policy resulted in final EV preference increases of 14% and 7% for *Innovation-oriented Progressives* (Fig. 5c) and *Eco-oriented Opinion leaders* (Fig. 5d), respectively, when compared to baseline. Tax exemption and purchase subsidy yielded up to 15% and 14% increase, respectively, in agents of type III, and 7% and 9% increase in agents of type IV. Mode shift in consumer type III mainly happened from the dominant ICE cars to EVs. In contrast, agents of mobility type IV reduced their preferences for the use of public transport and bicycle in favor of EVs, suggesting an ironic effect to a less sustainable form of transport choice in this consumer population. The variation of these fluctuation patterns across the four traveller groups shows that they are strongly constrained by the pre-existing mental representations and the following learning processes of agents in the simulation. The modeled policies are compatible with the consumer groups' initial representations to different degrees. Therefore, preference rates as well as their fluctuation differ across the scenarios, but remain constant beyond the simulated time scale, since the model is settled at time step 100.

Taken together, our simulation results contribute to current, political discussions about the anticipated consumer acceptance of EVs and about appropriate policy instruments to promote EVs. As shown in Fig. 6, the aggregated results across all agents yield optimistic projections for the acceptance of EVs under alternate sets of scenarios. In the reference case, EVs account for 15% of the total modal share. Simulations indicate that despite a considerably stable fraction of "early adopters" of EVs, social influence alone cannot effectively increase the spread of this innovation. Additional policy interventions are necessary to encourage a broader market penetration of EVs. The effects of simulated policy measures suggest a significant increase of potential adopters, with potential adoption rates converging over time up to 31% in the *Purchase subsidy*, 30% in *Zero-emission-zone* and 29% in *Tax exemption* scenario. Moreover, the results highlight in particular in short- and medium-term –indicated by the growth rate between time steps 1 to 35 in the *Zero-emission-zone* scenario– the role of non-financial policy strategies for increasing the acceptance rate and the potential adoption of EVs.

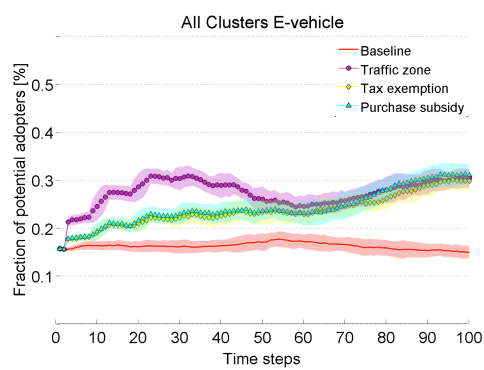

**Fig. 6.** Diffusion of E-Vehicles in *Reference case*, *Zero-emission-zone*, *Tax exemption* and *Purchase subsidy* scenarios across all agents

Although the results are explicitly not related to realistic time scales usually used in policy scenarios (i.e. months and years), the comparison of scenarios provides specific insights into the susceptibility of consumers to different policies and the consequent changes of mind. For example, our results provide additional evidence in a controversial discussion regarding the effectiveness of financial incentives on the purchase of EVs, showing that these subsidies do not generate additional consumer adoption beyond the existing tax rebates for EVs. This in line with previous findings (Diamond, 2009; Egbue & Long, 2012; X. Zhang, Wang, Hao, Fan, & Wei, 2013) and suggests the design of measures addressing non-monetary consumer needs such as independence and freedom from stress, as simulated in the traffic zone scenario.

## 4. Conclusions

We started with the premise that more psychologically realistic models of human decision-making are needed in innovation-diffusion simulations, in order to enhance their usefulness to political decision-makers and other practitioners. To address this need, we introduced a novel agent-based model (ABM), *InnoMind* (*Inno*vation diffusion driven by changing *Mind*s), based on state-of-the-art theories in cognitive science and grounded in empirical data.

We employed this framework to analyze three what-if scenarios representing different policy interventions–tax exemption, purchase subsidy and the introduction of an exclusive zone for EVs–intended to promote the adoption preference of electric vehicles (EVs). Our simulation results have three main implications. First, the failure to consider heterogeneous needs of different travellers reduces the effectiveness of the policies modeled in this paper in a significant manner. Second, somehow unexpectedly, the introduction of an exclusive zone for EVs in the city is in particular in the early phase of the diffusion process of EVs more effective than financial incentives only. This result emphasizes the importance to address non-monetary considerations of consumers in the development of policy measures. Third, mode switches to EVs induced by policies in consumer groups currently using bicycles as their main mode of transportation might, quite ironically, counteract the goals of $CO_2$ emission reduction.

In an additional baseline scenario where preferences of agents were only influenced by the interaction with their social peers, we did not find considerable shifts in transport mode preferences. This baseline simulation also suggests a relatively high and stable preference rate for EVs at least in two traveller groups among Berlin residents without additional interventions. Thus, we may conclude that travellers do not substantially change their pre-existing transport mode decisions when they interact with their social environment under constant external conditions.

As demonstrated above, the InnoMind framework offers the possibility to explore the emergent properties that result from rather complex individual decisions under combined influence of social dynamics and policy measures. The model allows one to explore how current empirically-derived preferences might change dynamically as a result of social communication. Considering the pivotal role of decision-making processes in agent-based models, our proposed model provides a novel view on how beliefs, needs, priorities and emotions drive individual decisions about transport mode use. Complementary to research in social psychology that aims to understand the influence of motives and intentions on modal choices (Anable, 2005; Bergstad et al., 2011; Steg, 2005), our agent-based model provides a pathway for exploring the dynamical nature of these mental representations. Moreover, in contrast to economically rational decisions -assuming a serial deliberative analysis of risks and expected utilities of alternative actions- our approach conceives of decisions as a result of an automatic holistic process, accepting a transport option merely if it maximally satisfies the constraints given by mental representations.

We belief that this concept of the human mind in combination with our proposed segmentation approach may contribute to guiding the development of more demand-side oriented policy instruments considering emotional as well as cognitive constraints of behavioral change when attempting to encourage more sustainable transport choices.

Finally, the flexible nature of our policy simulation system allows the implementation of a broad range of policy scenarios. Although conceptualized and calibrated in this work to study the potential diffusion of EVs, the model may also be adopted to other issues of social transition with modest effort.

## 5. Limitations and directions of future work

Although we are convinced of the practical relevance of our simulation results for decision-makers in politics and business, our approach has limitations to be addressed in future work. First, the showcase region Berlin, Germany, on which we focused in this study, is not representative for the travel preferences and behavior of Germans (BMVBS & DLR, 2008). Specific characteristics of the city –such as the urban environment, low rates of car ownership, well-developed public transport, and the innovative brand of Berlin– necessarily limit the impact of reported simulation results on nation-wide policy interventions. However, our framework could be easily expanded to further regions or even across Germany. Provided the availability of suitable empirical data, a national model of diffusion of transport innovation could be used for exploring geographically tailored policy strategies, in order to achieve a transition to a low-carbon transport system in the country.

Second, the model ignores the supply side. A comprehensive assessment of EV diffusion, however, requires an integrative approach including technological, political and societal influence factors (Tran et al., 2012). Different electric-drive technologies such as full battery, plug-in hybrid, and fuel-cell EVs, as well as new business models, transport services, and marketing campaigns of manufacturers will have considerable impact on market dynamics. Due to the lack of sufficient data in the start-up phase of EV innovations, we were not able to ground supply-side actors similarly to consumers' decision motives. In future research, we intend to extend our model to account for technological innovations and marketing campaigns of manufacturers and other suppliers, based on empirical data.

Third, we did not explicitly model varying perceptions of short-term and long-term costs of vehicle ownership. A related body of literature in behavioral economics provides evidence that people tend to prefer immediate payoffs than more distant ones in time (e.g., Laibson, 1997). In consideration of higher purchase prices but lower cost of maintenance for EVs relative to internal combustion engine cars, this is an important issue. At present, little is known about how much consumers are willing to pay for future fuel savings (for review, see Greene, 2010). In the present work, we addressed these inconsistencies in part by subjective weights of impacts in the two monetary policy scenarios (see Section 2.4). In future work, we plan to conduct behavioral experiments to even better inform our model empirically.

Despite these limitations, we believe that our agent-based model is a worthwhile research approach suitable for many further applications. Further activities could comprise the exploration of agency of households in transport innovation adoption or a more participative involvement of stakeholders (e.g., politicians or managers) in the modeling process (for review, see Barreteau et al., 2013). Moreover we intend to create a flexible, user-friendly surface of the software InnoMind to enable non-experts to design and simulate scenarios meeting their questions and assumptions.

**Acknowledgements**

This research was funded by the German Federal Ministry for Education and Research (BMBF) under grant #16|1610. Schröder's contributions were also supported by research fellowship #SCHR1282/1-1 from the German Research Foundation (DFG). We thank Paul Thagard for helpful discussions and comments on model development, Alexander Hatri for his contributions to programming the model in Java, Kerstin Stark and Jakob Dordevic for their assistance in data collection, and Jesse Hoey for valuable comments on an earlier version of the manuscript.

# Appendix

## Appendix A. Decision-Making of Individual Agents

The decision model HOTCO we used in this agent-based model combines previous connectionist models such as explanatory coherence (ECHO), analogical mapping (ACME), decision making (DECO) and impression formation (IMP) (Holyoak & Thagard, 1989; Kunda & Thagard, 1996; Thagard, 1992; Thagard & Millgram, 1995). Decision processes are represented by spreading activations and valences trough the network. In an iterative updating algorithm activations and valences of the units are updated until they reach a stable level, i.e., the network has settled. Activations are updated according the following equation:

$$a_j(t+1) = a_j(t)(1-d) + \{net_j[\max - a_j(t)]\} \; if \; net_j > 0$$
$$\{net_j[a_j(t) - min)]\} \; if \; net_j \leq 0, \tag{A.1}$$

where $net_j = \sum_i w_{ij} a_i(t) + \sum_i w_{ij} v_i(t) a_i(t)$ \hfill (A.2)

where $a_j(t)$ represents the activation of the unit $j$ at iteration $t$. The constant parameter $d$ (= 0.05) is the rate decay of activation for each unit at every cycle, $min$ is minimum activation (-1) and $max$ is the maximum activation. $Net_j$ is the valence and activation net input to a unit $j$ calculated based on the connection weight between unit $i$ and unit $j$ that is $w_{ij}$.

Valences of units are updated by a similar equation:

$$v_j(t+1) = v_j(t)(1-d) + \{net_j[\max - v_j(t)]\} \; if \; net_j > 0$$
$$\{net_j[v_j(t) - min)]\} \; if \; net_j \leq 0, \tag{A.3}$$

where $net_j = \sum_i w_{ij} v_i(t) a_i(t)$ \hfill (A.4)

where $v_j(t)$ represents the valence of the unit $j$ at iteration $t$. The constant parameter $d$ (= 0.05) is the rate decay of activation for each unit at every cycle, $min$ is minimum activation (-1) and $max$ is the maximum activation.

## Appendix B. Defining the Social Network of Agents

Social similarity $\Delta_{ij}$ between two agents $i$ and $j$ is defined as Euclidean distance $d$ in $m$ dimensions:

$$\Delta_{ij} = \sqrt{\sum_{m \in S} \left( \frac{S_{mi} - S_{mj}}{max\ d_m} \right)^2} \tag{B.1}$$

where $m$ is the number of socio-demographics factors; $S_m$ is the socio-demographic variable under consideration and $S_{mi}$ is the particular variable of an agent $i$. The similarity calculations are normalized by the maximum distance $d_m$ along that dimension that occurs within the agent population. Based on the similarity of two agents we defined a social tie weight $\delta_{ij}$ for each pair:

$$\delta_{ij} = 1 - \left( \frac{\Delta_{ij}}{max\ \Delta} \right) \tag{B.2}$$

The final probability that two agents form a social tie is additionally influenced by a stochastic factor $R$ ranging from 0 to 1. Formally expressed:

$$if\ R < \delta_{ij} \tag{B.3}$$

The resulting social network is static thus social ties do not change over the course of one simulation run.

# Appendix C. Initial Parameterization of Agents

**Table C1**: Input and output parameter settings

| Variable Type | Para-meter | Range or Value | Meaning | Input/ Output |
|---|---|---|---|---|
| ***HOTCO parameter*** Need units ($G_{1-8}$) | | | | |
| | $w_{ii}$ | -1 to +1 | Facilitation weight between need unit and action unit ≙ cognitive representation to which degree a need is accomplished by a certain action | Input |
| | $w_{ii}$ | -1 to +1 | Valence weight between need unit and special valence unit ≙ valence of a specific need | Input |
| | $p_i$ | -1 to +1 | Priority weight between need unit and special unit ≙ priority of a need | Input |
| | $a_i$ | -1 to +1 | Activation of a need unit ≙ impact of a need on decision making | Output |
| | $v_i$ | -1 to +1 | Valence of a need unit ≙ intrinsic emotional valence of a need or action | Output |
| Action units ($A_{1-5}$) | $w_{ij}$ | -1 to +1 | Facilitation weight between need unit and action unit ≙ cognitive representation to which degree a need is accomplished by a certain action | Input |
| | $w_{ij}$ | -1 to +1 | Valence weight between action unit and special valence unit ≙ valence of a specific action | Input |
| | $a_i$ | -1 to +1 | Activation of an action unit ≙ indicates which option to choose | Output |
| | $v_i$ | -1 to +1 | Valence of an action unit ≙ emotional valence of the associated mode of transport | Output |
| ***Socio-demographic/ - economic parameters*** | | | | |
| | $A$ | 18 to 69 | Age of an agent | Input |
| | $G$ | 0,1 | Gender of an agent | Input |
| | $I$ | 1 to 7 | Income categories | Input |
| | $E$ | 1 to 5 | Level of education | Input |
| | $C$ | 1 to 3.6 | Standard of consumption* | Input |
| | $M$ | 1 to 4 | Level of modernity* | Input |
| ***Further parameters*** | | | | |
| | $T$ | 1 to 4 | Different mobility types based on previous conducted cluster analysis | Input |
| | $r$ | 0 to 1 | Social radius ≙ social range of an agent limiting the size of the personal network | Input |
| | $X,Y$ | 0.33 to 0.68 | Geographic coordinates (latitude and longitude) assigned to postal codes | Input |
| | $\mu$ | 0 to 1 | Weight of policy influence | Input |

Note all input parameter are determined empirically based on our survey study.

**Table C2:** Mean (M) policy impact factors and standard deviations (SD) of different mobility types

| | Type 1 Comfort-oriented Individualists | | Type II Cost-Oriented Pragmatics | | Type III Innovation-Oriented Progressives | | Type IV Eco-Oriented Opinion Leaders | |
|---|---|---|---|---|---|---|---|---|
| | M | SD | M | SD | M | SD | M | SD |
| *Policy impact factor for* | | | | | | | | |
| Zero-emission-zone scenario | .45 | .25 | .48 | .22 | .60 | .20 | .63 | .21 |
| Tax exemption scenario | .47 | .26 | .52 | .27 | .66 | .22 | .71 | .22 |
| Purchase subsidy scenario | .53 | .28 | .56 | .26 | .69 | .23 | .71 | .23 |

## Appendix D. Modeling Persuasion in Agent-to-Agent Communication

The percentage values shown in Table D.1 and D.2 represent self-reported mean changes of opinions across subjects and transport modes (N = 480) as a result of positive and negative vignette statements using factual proposition (Table D.1) or emotional propositions (Table D.1). For instance, the first column in Table D.1 indicates that listener with a very strong preference for EVs ($w_{ij} \geq .60$) listening to a strong positive statement about EVs perceive a considerable enforcement of this belief (i.e. factor of rational influence $\pi = +8.3\%$). The influence of a negative statement, however, is for the same person almost negligible (i.e. factor of rational influence $\pi = -0.3\%$).

**Table D.1:** Percentage factors of rational influence $\pi$ for the information receiving agent in means-ends communication

| Senders facilitation weight $w_{ij}$ | Receivers facilitation weight $w_{ij}$ | | | | |
|---|---|---|---|---|---|
| | $w_{ij} \geq .60$ | $.20 \leq w_{ij} < .60$ | $-.20 \leq w_{ij} < .20$ | $-.60 < w_{ij} < -.20$ | $w_{ij} \leq -.60$ |
| $w_{ij} > .30$ | $\pi = +8.3\%$ | $\pi = +7.3\%$ | $\pi = +4.0\%$ | $\pi = -4.1\%$ | $\pi = -3.0\%$ |
| $w_{ij} < -.30$ | $\pi = -0.3\%$ | $\pi = -0.6\%$ | $\pi = -1.3\%$ | $\pi = -0.3\%$ | $\pi = -2.0\%$ |

**Table D.2:** Percentage factors of emotional influence $\alpha$ for the information receiving agent in contagion communication

| Senders valence weight $v_i$ | Receivers valence weight $v_{ij}$ | | | | |
|---|---|---|---|---|---|
| | $v_i \geq .60$ | $.20 \leq v_i < .60$ | $-.20 \leq v_i < .20$ | $-.60 < v_i < -.20$ | $v_i \leq -.60$ |
| $v_i > .10$ | $\alpha = +7.5\%$ | $\alpha = +3.5\%$ | $\alpha = +0.6\%$ | $\alpha = -1.0\%$ | $\alpha = -2.5\%$ |
| $v_i < -.10$ | $\alpha = +4.0\%$ | $\alpha = +0.35\%$ | $\alpha = -0.1\%$ | $\alpha = -0.85\%$ | $\alpha = -1.8\%$ |

The content of communication is selective. Therefore the speaking agent selects based on a threshold ($z$) for facilitation weights ($w_{ij}$) $z = \pm 0.3$ and for valences of actions ($v_i$) $z = \pm 0.1$, respectively, the content of the conversation.

Based on our experimental evidence we defined the percentage of weight change for means-ends communication mechanism as a function of strength of receivers' connection weights and the consistence of opinions. Incoming information of the speaking agent is therefore compared to the corresponding pre-existing belief structure of actions of the listening agent to select the according values for factor $\pi$ of rational influence (see Table D.1). Facilitation weights changes are formalized as

$$w_{ij}(t+1) = w_{ij}(t) \pm \Delta w_{ij} \qquad \text{(D.1)}$$

where

$$\Delta w_{ij} = \frac{w_{ij}(t)}{100 \times \pi} \qquad \text{(D.2)}$$

where $w_{ij}$ is the weight of the connection from unit $j$ to unit $i$, $t$ is the previous time step and $t+1$ represents the current time step.

Emotional influence in emotional contagion mechanism is implemented by adjusting special valence weights ($w_i$) of receivers' action units. Values of special valence weights ($w_i$) are set in each communication procedure based on receivers' current emotional connotation of an action ($v_i$) and the empirical determined factor of emotional influence $\alpha$. In contrast to means-ends communication the factor or emotional influence $\alpha$ is determined by the valence values of sender's and receiver's action units in question. Formally expressed by:

$$w_{ij}(t+1) = v_i(t+1) \pm \Delta w_{ij} \qquad \text{(D.3)}$$

where

$$\Delta w_{ij} = \frac{w_{ij}(t)}{100 \times \alpha} \qquad \text{(D.4)}$$

where $w_{ij}$ is the weight of the connection from special valence unit $j$ to action unit $i$ and $t+1$ represents the current time step. Note that values for special valence weights are not accumulated over time assuming that agents are merely affected by the emotional input of their current conversation partner not by previous discussions.

**Appendix E. The Media Agent**

The influence of media agent on mental representations of agents is calculated similar to the communications mechanisms described in Appendix D. The weight update is formalized:

$$w_{ij}(t+1) = w_{ij}(t) \times \Delta w_{ij} \qquad (E.1)$$

Where

$$\Delta w_{ij} = \mu \qquad (E.2)$$

The factor $\mu$ represents the evaluation of policy impact on an individual level, derived from ratings in our online survey. In all policy intervention simulated in this work $\mu$ was based on survey responses. For technical reasons we transformed the original response on a five-point Likert scale ranging from 1 (absolutely no influence) to 5 (very strong influence) to a parameter range from 0 to 1.

## Appendix F. Sensitivity analysis

We conduct a sensitivity analysis on the policy impact factor $\mu$ using five different parameter settings ($\mu$ = empirical/.25/.50/.75/1.00). We conducted one model runs for each simulation to assess the effect on the preferences for EVs across the hole agent population for each setting. We look at the effects on EV preferences at time step $t = 100$.

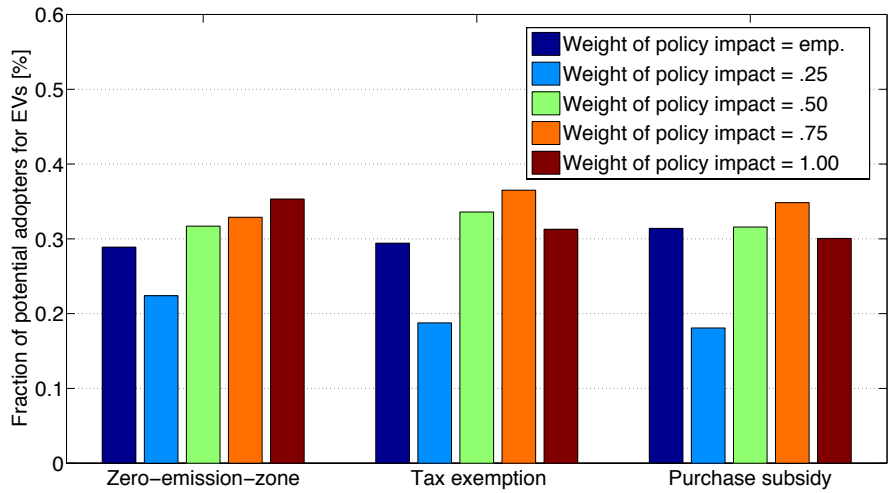

Fig. F1. Preferences for EVs at time step $t = 100$.